# FIRST COBOTIC-ASSISTED STRING ASSEMBLY OF PIP-II SSR2 CAVITIES AT FERMILAB*

M. Parise†, T. Aiazzi, C. Narug
Fermi National Accelerator Laboratory, Batavia, IL, USA

*Abstract*

Achieving robotic-assisted string assembly has been a longstanding objective in superconducting cavity production to improve precision, repeatability, and throughput while reducing operator-dependent variability. At Fermilab, robotic systems have been successfully deployed for the installation of individual components on SSR2 cavities inside a cleanroom environment. The single-cavity installations validated the feasibility of robotic techniques and provided confidence to extend automation to the assembly of a full prototype cavity string. Leveraging lessons learned from earlier operations, the robotic system was adapted for the string assembly. This work describes the robotic-assisted assembly process in detail, including alignment strategies for coupler flanges, torque control for bellows connections, and mitigation of particulate generation in cleanroom conditions. Performance metrics of the cryomodule are presented, together with an analysis of improvements needed for robotic assembly. The effort also represents a broader R&D initiative at Fermilab to the long-term goal of scalable SRF cryomodule production.

## INTRODUCTION

The Proton Improvement Plan-II (PIP-II [1]) project at Fermilab entails the procurement and assembly of a large number of superconducting radio-frequency (SRF) cryomodules. The spoke cavities section alone comprises of sixteen Single Spoke Resonator type 1 (SSR1) [2] and thirty-five Single Spoke Resonator type 2 (SSR2) [3-6] jacketed cavities, distributed over the two SSR1 [7] and seven SSR2 [8] cryomodules. Every cavity to be integrated into a cavity string along with superconducting magnets, Beam Position Monitors (BPMs), bellows and gate valves and each string shall be assembled without contaminating the beam volume.

SRF cavity strings are assembled in cleanrooms, generally of ISO class 4 or 5 according to ISO 14644-1 [9]. Microscopic particulate contamination on the interior surfaces of a cavity may trigger field emission, may cause local quenching and may degrade both the quality factor and the accelerating gradient. Each individual component is therefore wet-cleaned with specialized detergents in ultrasonic baths, the cavities receive high-pressure rinsing (HPR), the quality and the effectiveness of which can now be predicted by simulation before the process is executed in the cleanroom [10], and filtered air is pushed continuously from the ceiling towards the floor or the sides of the room, so that any particulate generated during the operations is swept away from the assembly area.

Since all parts are thoroughly cleaned before they enter the room, the operator becomes the predominant source of contamination during any cleanroom assembly. The skills and the experience of the cleanroom operators, along with the design of the tooling, are paramount for a successful string assembly [11, 12].

Manual tooling for cleanroom assembly is a double-edged sword. A typical device requires six degrees of freedoms (DOFs) to regulate the position of the component being installed and the operator is not the one holding the weight, as shown in Fig. 1. It must be flexible and generate the least possible amount of particulate and comply with cleanroom design principles, with sliding connections and threaded features placed at the bottom whenever possible, away from the assembly area. A six DOFs machine, however, is never truly particle-free, nor easy to use: it requires training, dry fits and mock-ups. Finally, as soon as the component is revised, slightly changed, or the next project starts, custom tooling machined from stainless steel and anodized aluminium becomes obsolete, and it is expensive.

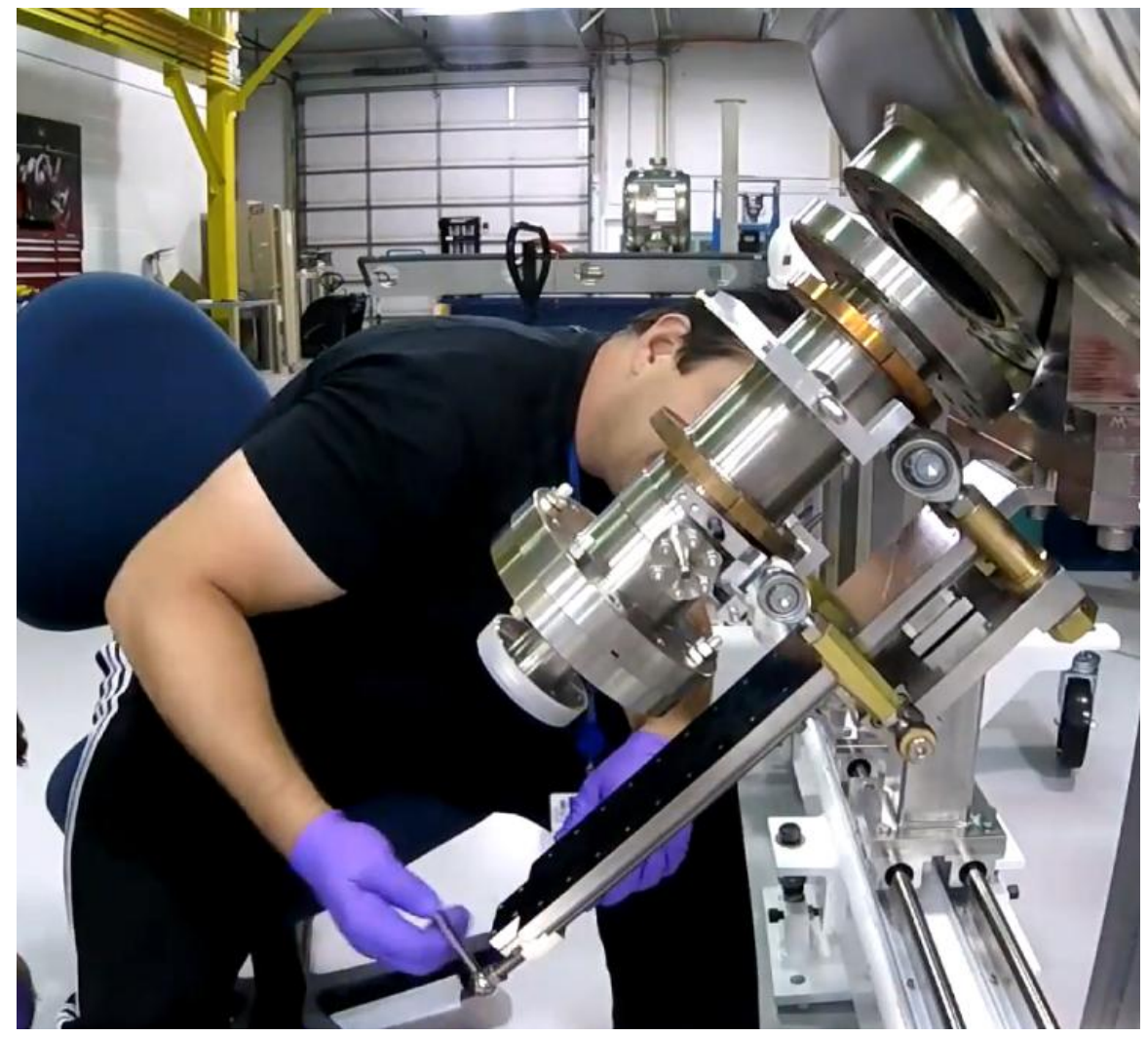

Figure 1: Manual, tooling-assisted installation of a component onto an SRF cavity in the cleanroom.

## THE ROAD TO THE FIRST ASSEMBLY

In late 2023 Fermilab procured its first collaborative robot, a Universal Robots UR16e [13], with the goal of developing cobotic applications for the cleanroom assembly of SRF components, specifically for PIP-II. The main characteristics of the arm are summarized in Table 1. Two additional cobots, one UR16e and one UR5e, have been procured since.

† mparise@fnal.gov

Three principles were established at the beginning. First, open-source options were to be used whenever possible, rather than plug-and-play commercial solutions. This costs development time, but it was considered essential in a field where every component is custom made, identical assemblies are rarely repeated (and proton machines feature many different cavity types) and strings often feature superconducting magnets and BPMs. Secondly, collaborators and students should be included whenever possible. Third, the initial objective had to be simple: to align, approach and install a vacuum-end high-power coupler.

Table 1: Main Characteristics of the UR16e Cobot

| Feature | Value |
|---|---|
| Axes | 6 |
| Payload capacity | 16 kg |
| Reach | 900 mm |
| Integrated sensors | End-effector load cell |
| Force accuracy | ±5.5 N |
| Torque accuracy | ±0.5 N m |
| Cleanroom rating | ISO class 5 at ≤ 40% speed |
| Control system | PolyScope 5 pendant |

Since it was not known which strategy would suit the application, a total of four alignment approaches were developed in parallel: purely visual alignment, a camera system, alignment pins on the coupler side and threaded rods on the cavity side. A development test station was set up on the production floor, where damaged and 3D-printed components were used to build a first validation model and to prepare the cleanroom setups.

Within four months the team went from unpacking the cobot to a complete cleanroom setup, which required a dedicated cart as well as new procedures and safety protocols. The first cobotic installation of a vacuum-end high-power coupler onto an SSR2 cavity was performed in December 2023 [14] and it is shown in Fig. 2.

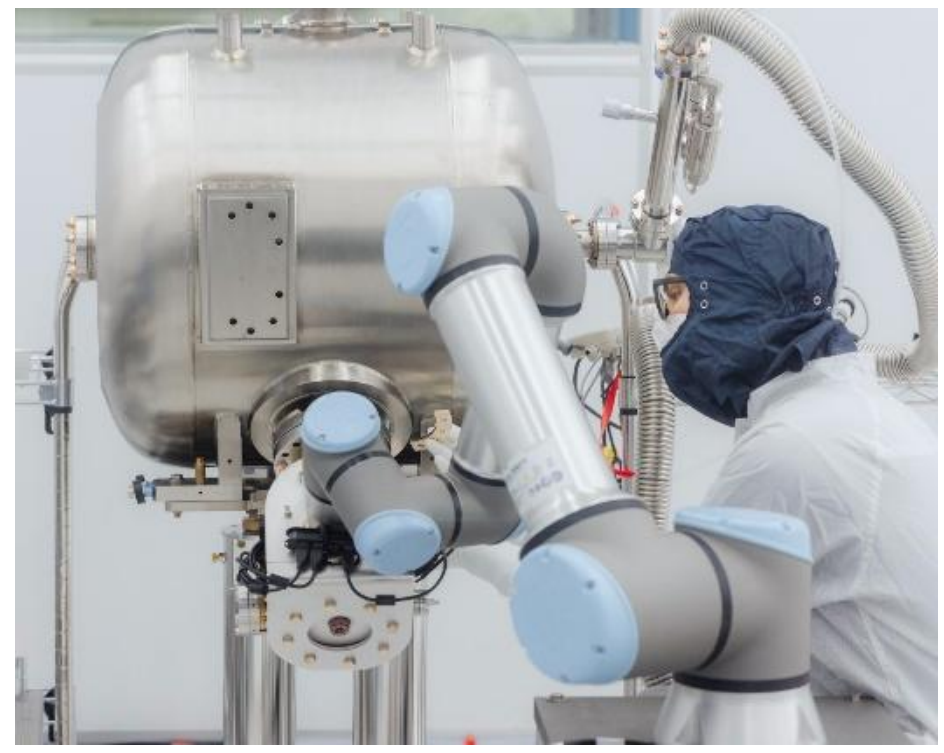

Figure 2: First cobotic installation of a vacuum-end high-power coupler onto an SSR2 cavity, December 2023.

The procedure was divided into steps: setup of the room; positioning of the arm at a pose that allows the operators to blow the vacuum end coupler clean; pre-assembly setup, performed with the cavity still closed, during which pins and cameras are installed on the coupler; alignment; blow cleaning of the coupler; removal of the cavity flange; installation; sealing of the coupler to the cavity flange; and installation of the bolts. The operators stay as far away as possible while the cavity flange is removed and the beamline volume is exposed to the cleanroom environment. They come back only to install the bolts, at which point the risk of contamination is severely reduced, because the cobot actively pushes the coupler onto the cavity flange and attempt to seal out any particulate that may be generated.

## THE PRE-PRODUCTION SSR2 STRING ASSEMBLY

Three more couplers and several beamline bellows were subsequently installed on single SSR2 cavities with demonstrated success (the cavities that were affected by field emission did not show any degradation and those that were field emission free, remained as is). These installations validated the technique and gave enough confidence to use the cobot for most of the components of the pre-production SSR2 (ppSSR2) cavity string. The composition of the string is given in Table 2. The jacketed cavities had previously been qualified, some with high power couplers and tuners, in the Fermilab Spoke Test Cryostat [15].

Table 2: Composition of the ppSSR2 Cavity String

| Component | Quantity |
|---|---|
| Jacketed cavities | 5 |
| Magnets and BPMs | 3 |
| Interconnecting bellows | 9 |
| Gate valves | 2 |

To prepare for the string assembly, additional single cavity assemblies were performed on the development test station outside the cleanroom. Opportunities for improvement were identified, excess equipment was removed, the alignment pins were selected for the final process and the bellows tooling was developed. The cobot was used for the installation and the alignment of the nine interconnecting bellows, shown in Fig. 3, as well as of three vacuum-end couplers.

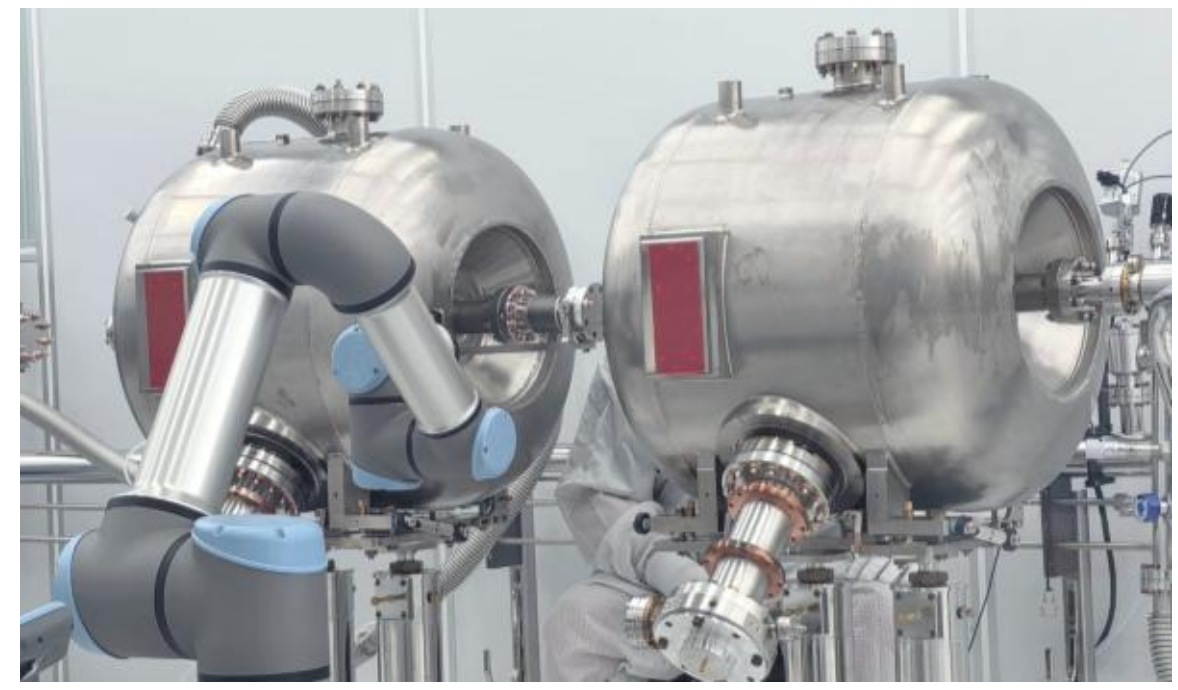

Figure 3: Cobotic alignment and installation of one of the nine interconnecting bellows of the ppSSR2 cavity string.

The assembly started in September 2025 and the string, shown in Fig. 4, was completed in October 2025, over approximately four weeks. It was integrated into the ppSSR2 cryomodule [8].

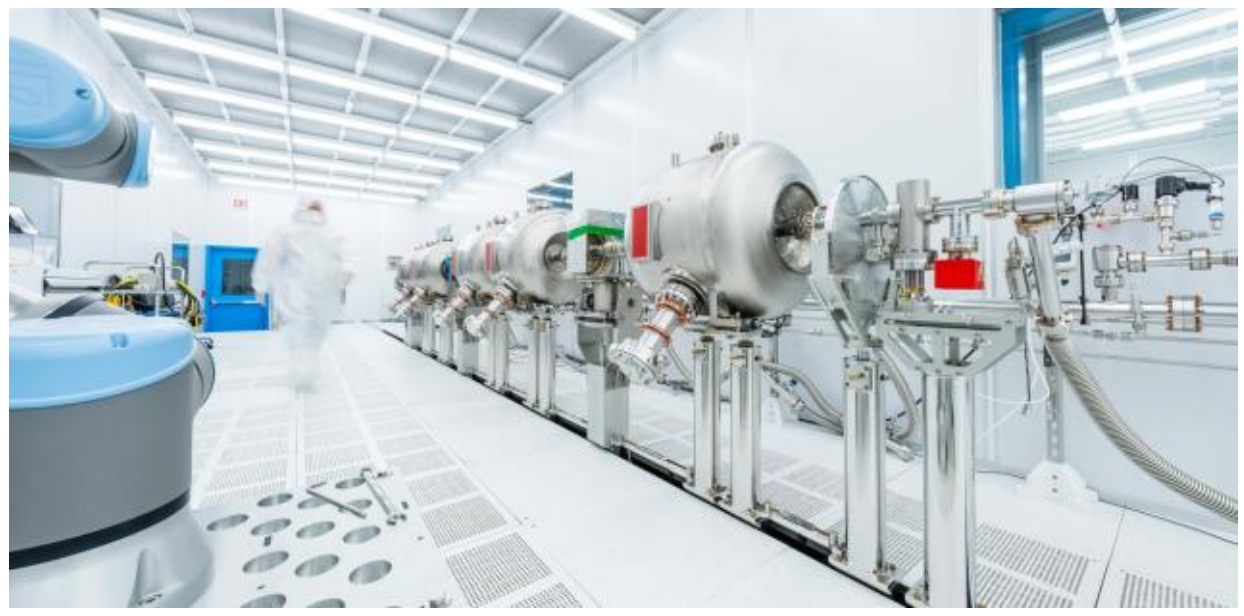

Figure 4: The pre-production SSR2 cavity string at the end of the first cobotic-assisted string assembly, October 2025.

Overall, during string assembly the limits of the approach become more evident: the alignment still relied on the visual feedback of an operator, who must remain very close to the open cavity/magnet/bellows. Moving the operator one step away, and thereby further reducing the contamination risk, became the objective of the next development phase.

## THE NEXT PHASE: FULL AUTOMATION

The next challenge was a fully automatic assembly, with no operator involved in the alignment. The operator remains in the room but stays far from the parts being assembled. The task selected was the one the team had started with three years earlier: the assembly of a vacuum-end coupler. This time on an SSR1 cavity [2, 7], with a similar setup into which professional cameras were integrated.

### *Hardware*

The UR16e carries the coupler on its bracket; the 16 kg payload accommodates the coupler and the bracket as well as the cameras. An Intel RealSense D435i , and a Basler a2A4508 18-megapixel monochrome camera with a 16 mm Fujinon lens are mounted next to each other on the wrist and used in dual-camera mode; the Basler is a fixed-focus, high-resolution device. Force sensing is provided by the cobot itself, which measures the force applied at the end effector. The setup is shown in Fig. 5.

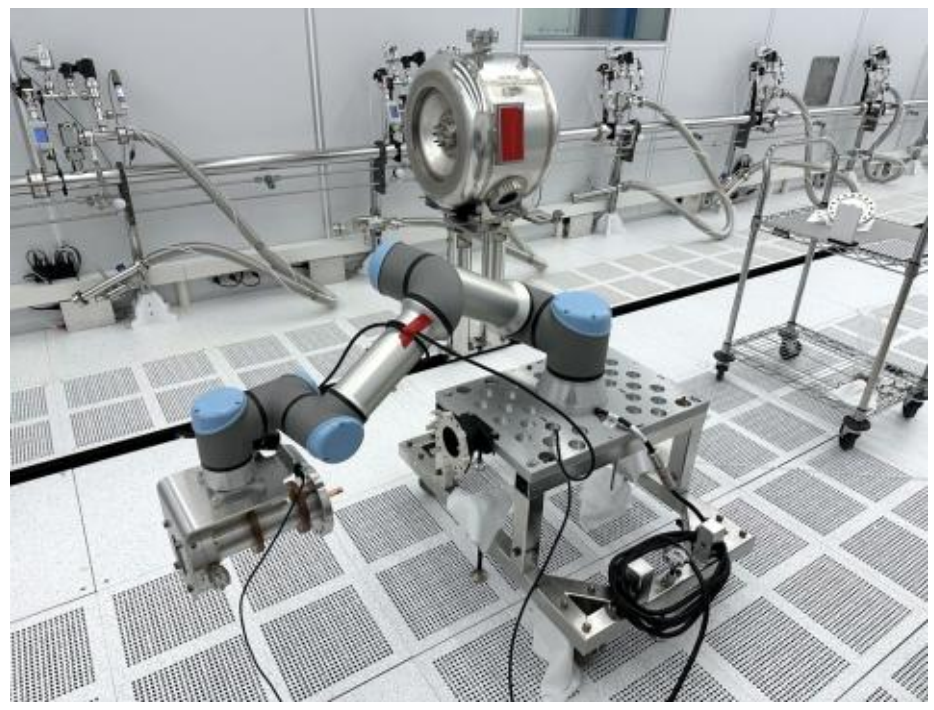

Figure 5: Setup for the fully automated assembly: the UR16e is mounted on a cart and carries the vacuum-end coupler and the two cameras.

### *Camera Calibration*

The pipeline mainly consists of three steps. The first, camera calibration, provides the camera intrinsics, that is the distortion of the optics, and the extrinsics, that is the position of the camera with respect to the end effector, also known as hand-eye calibration. A dedicated program automatically centres the camera frame on a fixed checkerboard and moves the cobot through 54 poses; at each pose an image of the checkerboard is acquired and used to compute intrinsics and extrinsics. The procedure lasts about 5 min and is required only the first time: as long as the camera is not touched, there is no need to repeat it at the next assembly.

The 54 poses are complex in nature and the process runs unattended, so a second program was developed that executes the full calibration sequence in NVIDIA Isaac Sim [16] before it is run in the cleanroom, as shown in Fig. 6. It reports whether any of the poses would generate a collision and whether all the paths are plannable. It also indicates where the checkerboard should be placed to obtain a 100% success rate, and how much the camera can be tilted and rolled, larger angles giving better results.

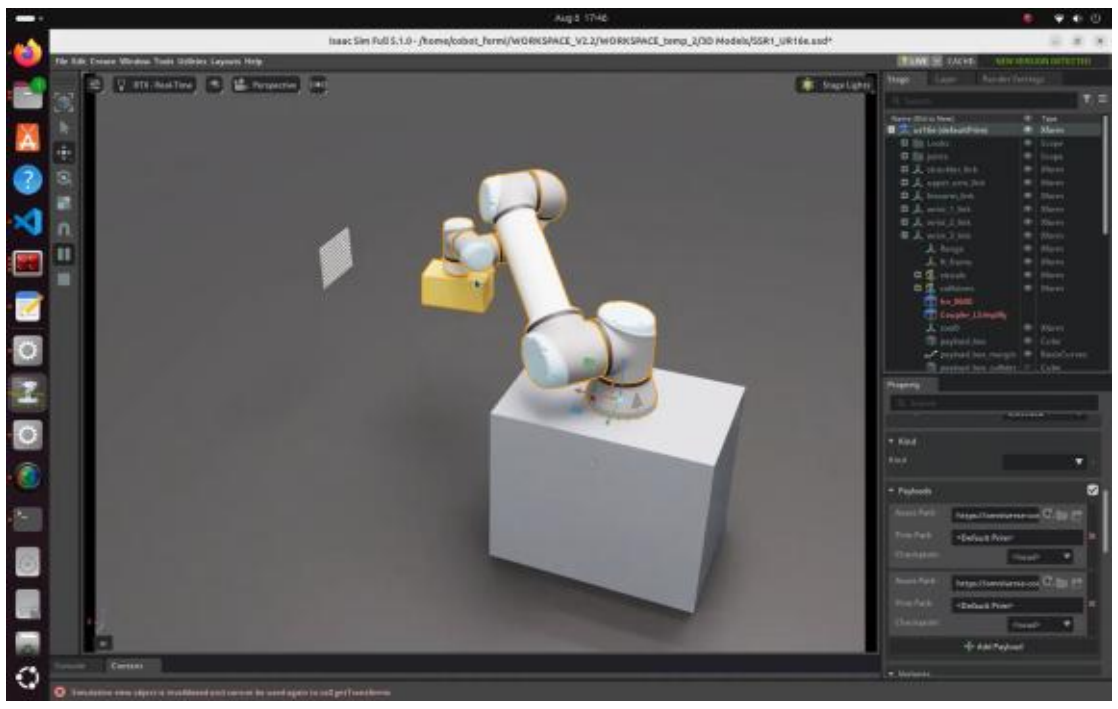

Figure 6: Simulation in NVIDIA Isaac Sim of the 54 hand-eye calibration poses, used to verify reachability and to exclude collisions.

### *Coupler Calibration*

The second step answers the question of where the coupler is with respect to the end effector. A fixed camera looks at the coupler flange while the cobot moves the coupler through a set of known poses; by detecting the centre of the flange and of its holes and solving the system $AX = YB$ over all the poses, minimising the residuals, the position of the coupler with respect to the end effector is obtained in about 5 min. Subsequently, to ensure the calibration was successful the camera detects a fixed calibration flange, the cobot aligns the coupler to it and the pose may be fine-tuned at this stage. A plastic flange is used to verify it before moving on to the cavity, after which the coupler is blown clean and is ready for installation. Unlike the camera calibration, the first part of this step has to be repeated at every installation with the final checking as optional.

### *Markerless Detection*

The detection of the cavity flange is markerless and relies on two locally executed AI models. OWLv2 [17], a zero-shot open-vocabulary detector, finds the flange and crops the image, and then selects the holes inside the crop; a simple gate on size and aspect ratio filters out unwanted features. SAM 2 [18], which outlines and tracks objects in images and videos, is then used to create a mask that excludes everything but the bolt holes. Ellipses are fitted to the segmented regions and their centre points are computed on the image, as shown in Fig. 7. The centers feed a Perspective-n-Point (PnP) solver, which returns the pose of the flange in the camera frame. The process works remarkably well: 15 to 16 holes out of 16 are located in every frame, and the reprojection error is of the order of one pixel.

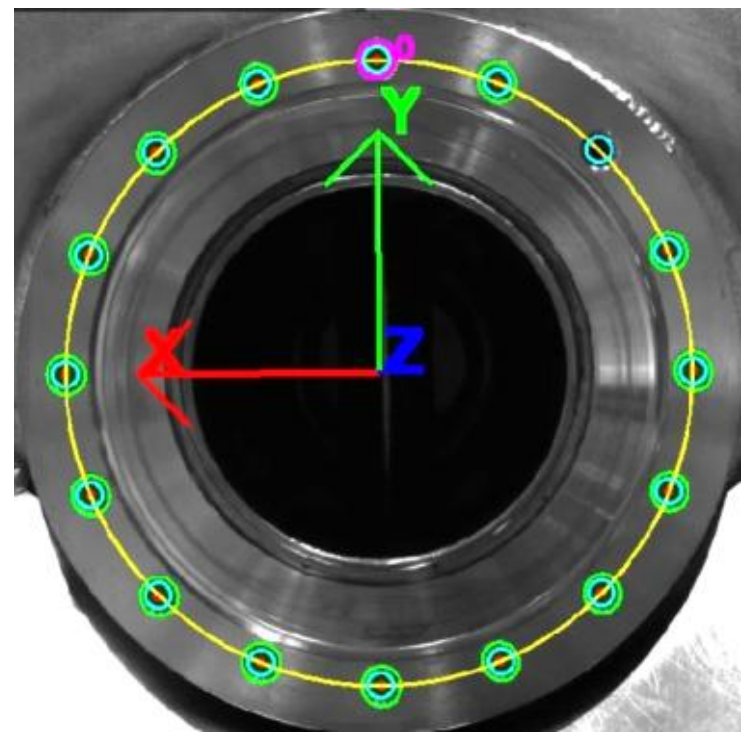

Figure 7: Markerless detection of the cavity flange. The bolt holes are segmented and fitted with ellipses, and their centres are used to solve the PnP problem.

Knowing where the camera and the coupler are with respect to the end effector, and where the cavity flange is with respect to the camera, the motion required to mate the coupler with the cavity flange is fully determined.

### *Multi-View Fusion*

A single image taken from a single point of view is not sensitive enough to compute the pose of the cavity flange with sub-millimetre and sub-degree accuracy. Multi-view image fusion was therefore exploited: the camera is moved around the detected flange and the estimated poses are fused, the rotational part being combined with the quaternion averaging algorithm of Markley et al. [19]. The scheme also keeps the system complexity low, since one camera is sufficient.

### *Main Assembly*

The main assembly lasts about one minute depending on the speed selected for the movement of the cobot, excluding the manual installation of the fasteners, and proceeds through the following steps:

- detection and initial accumulation of the poses of the cavity flange;
- coaxial alignment of the camera, with the cavity flange centred on the camera frame at a distance and left to settle for 1 s to 3 s;
- frontal accumulation of five samples, followed by four oblique orbits of ±10° to ±20° about the flange X and Y axes, with automatic fallback to ±5° if a view sees fewer than ten holes;
- fusion of up to five estimates into the final target, with divergence gates that stop the sequence in the presence of a systematic error;
- final approach with the pose frozen, from 200 mm to 100 mm and then to 0 mm, in straight-line Cartesian motions with one plan-and-confirm gate per step;
- force seating: while the measured axial force is below the set value, in the range 20 N to 100 N, the coupler advances in steps of 0.5 mm along the target axis, up to 10 mm past the estimated plane, after which the sequence stops with a warning.

### *Extension to Other Cavity Types*

The same pipeline was recently applied, with success, to a high-beta 650 MHz (HB650) cavity for PIP-II, shown in Fig. 8. On this particular coupler-to-cavity interface the gap between the bolts and the holes is about 0.4 mm over twelve holes, which makes the process considerably more challenging than on the spoke cavities.

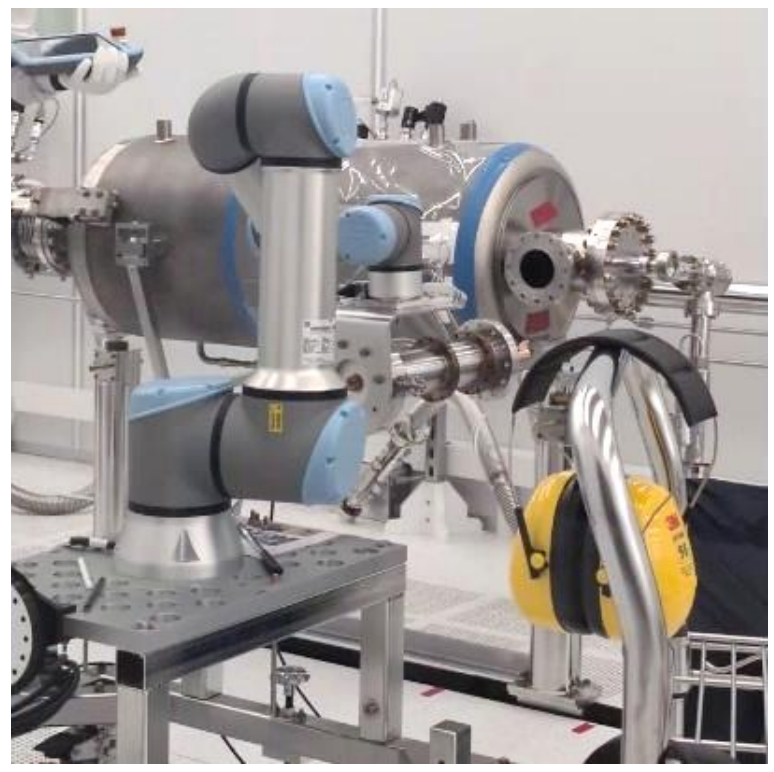

Figure 8: Fully automated coupler installation on a PIP-II HB650 cavity.

## AUTOMATED HARDWARE CLEANING

Before cavities can be assembled in the cleanroom, all the hardware has to be cleaned. The process is simple but time-consuming, and most of its variability comes from human error, which makes it an attractive candidate for automation. A prototype system uses a second cobot, a UR5e, to blow the hardware clean automatically, as shown in Fig. 9. The main limit to automation was reading the particle counter: a custom control box was built for this purpose, so that the cobot receives the feedback from the particle counter and knows when a part is clean. Parts have been successfully cleaned in the test setup.

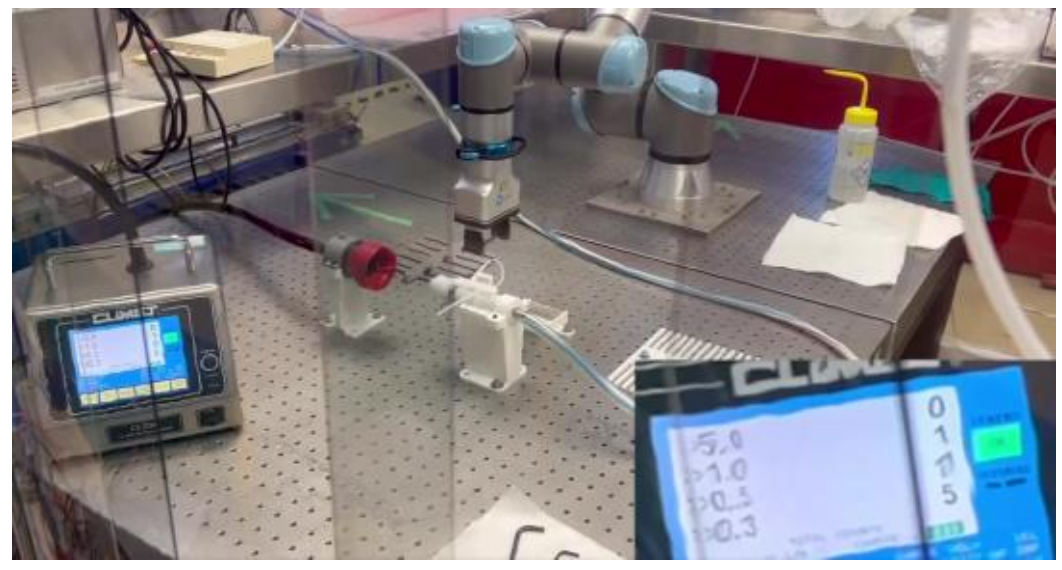

Figure 9: Prototype system for the automated cleaning of hardware, driven by the feedback of the particle counter.

# CONCLUSION

Leveraging three years of platform development, a markerless coupler-to-cavity alignment and assembly system has been developed that eliminates physical targets on the cavity, and the first cobotic-assisted assembly of a complete SRF cavity string, the pre-production SSR2 string for PIP-II, was completed in October 2025. All the processes and techniques presented here have been replicated in NVIDIA Isaac Sim, which opens the way to the deployment of AI training models to further optimize and refine the assembly procedures.

Because the automation architecture is developed in-house and tailored to the task, rather than bought as a turnkey industrial product, it can be adapted and redeployed to any high-precision operation requiring cobotically assisted inspection, ultra-clean assembly or automated quality control. Work continues on the extension of the fully automated process to the remaining components of a cavity string and on the reduction of the residual manual steps, with the long-term goal of a scalable and reproducible SRF cryomodule production.

# ACKNOWLEDGEMENTS

The work presented here is the result of a large collaboration between many groups. The authors would like to thank D. Passarelli, G. Wu and J. Bernardini; the cleanroom team and the machinists M. Battistoni, D. Bice, M. Chlebek, T. Fermanich, R. Kirschbaum, T. Ring and D. Santucci; and the university partners A. Ciaramella and A. Vivai (University of Pisa) and C. Denton, N. Giffen, J. Imburgia, J. Ryu, I. Salehinia and B. Whitlock (Northern Illinois University).